\documentstyle[12pt,aasms4]{article}
\begin{document}
\title{\bf The Variability of Active Galactic Nuclei and the Radial
Transport of Vertical Magnetic Flux}

\author{Seok Jae Park}

\affil{Korea Astronomy Observatory, San 36-1, Hwaam-Dong Yuseong-Gu,
Daejon 305-348, Korea I: sjpark@apissa.issa.re.kr}

\and

\author{Ethan T. Vishniac}

\affil{Department of Astronomy, University of Texas, Austin, TX 78712
I:ethan@astro.as.utexas.edu}

\begin{abstract}
We consider the radial buoyancy of vertical magnetic field lines in
radiation and gas pressure dominated accretion disks.  We find that
in addition to radial drift driven by turbulent diffusion and biased
by the global field geometry,
there are buoyancy effects which tend to move magnetic flux outward.
In gas pressure dominated disks the poloidal magnetic
field will move outward at a rate comparable to its inward advection.
On the other hand, in a radiation pressure dominated disk
the poloidal magnetic field will usually move outward faster
than it is advected inward.  This implies that the fields in disks in
active galactic nuclei are generated at small radii by an internal disk
dynamo.  This conclusion can be avoided if the external field imposes
a supersonic Alfv\'en speed within the disk without giving rise to
interchange instabilities.  In any case we note that
variations in the mass transfer rate will lead directly to a modulation
of the nonthermal emission from the disk system.
\end{abstract}

\section{Introduction}
It is well known that the
emission-line spectra of active galactic nuclei (hereafter AGNs) are
strongly variable.
Typical time scales for this variation range continuously
from the time scales appropriate for light travel across the inner
regions of the accretion disk around a black hole, i.e.
\begin{equation}
t\approx GM/c^3,
\label{eq:1p1}
\end{equation}
(hours for supermassive black holes) up to years.
This variability remains one of the central issues in any theoretical
model of the central engine in an AGN.

A popular model for the source of the nonthermal radiation from an
active galactic nucleus (AGN)
consists of a supermassive black hole surrounded by
a magnetized accretion disk.  In its simplest form this model
is taken to be axisymmetric and stationary.  The nonthermal
radiation is ascribed to
the Blandford-Znajek process (Blandford and Znajek 1977),
which emits nonthermal radiation from an extended magnetic field
embedded in the black hole event horizon, the surrounding plasma,
and the accretion disk.
This model
was reformulated and extended by
Thorne and Macdonald (1982), Macdonald and Thorne (1982),
and Thorne et al. (1986) in the `3+1'-spacetime formalism.
The magnetic field in the disk will
also give rise to a magnetically driven wind, and possibly
a jet, as matter spirals out along the poloidal field lines
(cf. \cite{bp82}, \cite{NS94}, \cite{SNRL94}, and \cite{SNORWL94}).
In addition, the ejection of toroidal flux from the disk may lead to
winds and jets (e.g. \cite{C95}).

Starting from the axisymmetric, stationary model for emission
due to the Blandford-Znajek process
Park and Vishniac (1989$a$, 1989$b$)
explored a time-dependent model which included the effects of variations
in the mass accretion rate.
The main point was to add the secular effects of mass accretion
to the original axisymmetric, stationary model of the nonthermal
radiation. Such variations are expected because
AGN accretion disks
are expected to be radiation pressure and electron scattering dominated,
and this can give rise to thermal and viscous instabilities on a broad
range of time scales (\cite{le74}).
In these two papers  we investigated the axisymmetric,
nonstationary
electrodynamics of a black hole and its accretion disk, respectively.

In the latter paper we found that the
electrodynamic power output from the accretion disk can be variable
on time scales associated with secular disk instabilities.
The point was that the local fluctuations in fluid velocities
in the accretion disk will cause fluctuations in the nonthermal
component of the radiation by the Blandford-Znajek process.
The time scales for these fluctuations, therefore, will reflect the
range of orbital periods in the inner annulus of the disk.

In this paper we discuss another effect which may explain the
fluctuations in the nonthermal emission, the relative motion
of the infalling matter and the entrained poloidal field. The traditional
view of the magnetic field in astrophysical objects, like accretion
disks, is that the matter and magnetic field lines are tightly
anchored to one another.  Consequently accretion disks should
continuously accrete poloidal field from the surrounding
medium.  Since the back reaction from the field on the matter
flow will also increase without limit this leads to a situation
in which the magnetic field completely overwhelms the gas pressure
and drastically alters the properties of the accretion disk.

On the other hand, if one supposes that the magnetic field is
not tightly coupled to the underlying fluid, then one can
invoke the concept of turbulent diffusion for the magnetic
field.  This leads to the prediction that large
scale magnetic fields tend to move outward in accretion disks
and are eventually ejected from their outer edges (\cite{vb89}).
Although it is difficult to reconcile turbulent diffusion with
the view that the magnetic field fills the entire fluid volume,
it is consistent with the idea that the magnetic field in stars and
accretion disks naturally forms intense, partially evacuated, fibrils
that are capable of moving relative to the bulk of the fluid, i.e.
flux tubes.
In a recent paper Vishniac (1995a) examined the underlying physics
behind the formation of such flux tubes and compared the predictions
of a particular model with numerical simulations.  A subsequent
paper (Vishniac 1995b) predicted that the flux tubes in ionized accretion
disks will be largely empty and discussed the consequences
of this model for the vertical buoyancy of the toroidal field
of an accretion disk.  Here we extend this work to the radial
motion of a poloidal field and show that in a radiation pressure
dominated disk it leads to a strong outward motion of the field
lines.  This in turn leads to a modulation of the nonthermal power
emitted by the magnetosphere as changes in the state of the disk
raise and lower the rate at which the poloidal field is dispersed
outward.

In \S II of this paper we discuss the radial buoyancy of
vertical magnetic field lines embedded in an accretion disk.
In \S III we discuss some of the implications of
this work for nonthermal radiation from AGN.

\section{The Radial Buoyancy of $B_z$}

In order to understand the transport of flux within an accretion
disk we need to have some model for the turbulent motions with
the disk.  Our work here is based on the idea that in an accretion
disk these motions are driven by the magnetic field with the result
that the angular momentum transport takes place within the disk and
is mediated both by hydrodynamic transport and shearing of the
local magnetic field.
A magnetized accretion disk will be subject to a violent shearing
instability which will act to transport angular momentum outward
(\cite{v59}, \cite{c60}, \cite{bh91}).  The corresponding dimensionless
viscosity $\alpha$ will be roughly the ratio of magnetic to thermal
energies or
\begin{equation}
\alpha\sim {B^2\over 8\pi P}\sim\left({V_A\over c_s}\right)^2,
\label{eq:alpha}
\end{equation}
where $B$ and $P$ are the characteristic values of the magnetic field
and pressure, respectively, in the disk, and $V_A$ and $c_s$ are the
corresponding values of the Alfv\'en speed and sound speed.
The resulting turbulence is  anisotropic, but only by factors of
order unity.  This turbulence will be characterized
by an eddy size of order $V_A/\Omega$ and a typical turbulent
velocity of order $V_A$.  Larger scale modes, as well as the
Parker instability, will be largely suppressed by these eddies
(\cite{vd92}).  Nevertheless, there will be a residual mean vertical
buoyant velocity of order $\alpha c_s$.  A similar result applies when
the magnetic field is considered to consist of individual flux tubes instead of
some diffuse, space-filling mean field (\cite{v95b}).
{}For a thin disk the sound speed $c_s\sim H\Omega$,
where $H$ is
the disk thickness and $\Omega$ is the rotational frequency.
Consequently, the dominant turbulent eddy size will be
\begin{equation}
L_T\sim \alpha^{1/2} H.
\end{equation}

The magnetic field in a turbulent medium will tend to separate into
regions of high and low magnetic field density.  Most of the magnetic
energy will be contained in flux tubes whose internal density is
kept at a level well below that of the ambient medium through turbulent
pumping (\cite{v95a}).  These flux tubes will interact with the surrounding
fluid through the turbulent drag of the fluid as it moves by.  Their
mutual interactions guarantee that they will be broadly distributed
in the fluid.  The force
per unit length on a flux tube is
\begin{equation}
\rho_{t}{d^2\vec V_{t}\over dt^2}(\pi r_t^2)=C_d\rho |\vec V_T-\vec V_{t}|
(\vec V_T-\vec  V_{t})r_t
+ \hat n_c (\pi r_t^2){B_{t}^2\over 4\pi R_c}-(\rho-\rho_{t})\vec g
(\pi r_t^2),
\label{eq:force}
\end{equation}
where the subscript $t$ denotes quantities evaluated within the flux
tubes, $R_c$ is the radius of curvature of the local magnetic field,
$\vec V_T$ refers to the local fluid velocity, $C_d$ is the coefficient of
turbulent drag (of order unity), $\vec g$ is the local gravitational
acceleration, and $\hat n_c$ is a unit vector pointing in the direction
of the curvature of the flux tube.  We will subsequently use
$|V_T|$  to refer to the root mean square turbulent velocity in the
fluid.
Only the velocity components perpendicular to the local magnetic field
enter into equation (\ref{eq:force}).  If the flux tubes are largely
empty, as they are in highly ionized disks, then the right hand side of
this equation must sum to zero. Under most circumstances the buoyant
force term will be much smaller than the other two terms (\cite{v95b}).
The properties of the flux tubes are largely
independent of the resistivity of the gas provided that
\begin{equation}
{\cal M}_T^4 \left({|V_T|\over k_T\eta}\right)\left({4E_B\over f_cw
E_T}\right)^{1/n+5/2}
>\left({4\over C_d\gamma}\right)^2,
\label{eq:diffcrit}
\end{equation}
where ${\cal M}_T$ is the turbulent Mach number, $k_T$ is the wave
number of the large scale eddies, $\eta$ is the resistivity, $E_B$
and $E_T$ are, respectively, the spatially averaged magnetic and kinetic
energy densities, $f_c$, and $w$ are constants of order unity,
$\gamma$ is the adiabatic index of the fluid, and
$n$ is the turbulent power spectrum index (equal to $2/3$ for
Kolmogorov turbulence).
The viscosity can be ignored if
\begin{equation}
1< \left({C_d\gamma\over 4\pi^3}\right){\cal M}_T^2
\left({|V_T|\over k_T\nu}\right)
\left({4E_B\over f_c wE_T}\right)^{3/2+1/n}.
\label{eq:diffvis}
\end{equation}
These conditions define the ideal fluid regime, within which the
flux tubes are largely evacuated, except for a surface layer whose width is
proportional to the square root of the resistivity.
In accretion disks the kinetic and magnetic energy densities are roughly
comparable due to the Balbus-Hawley instability.
In this case, equations (\ref{eq:diffcrit}) and (\ref{eq:diffvis}) can be
rewritten as (\cite{v95b})
\begin{equation}
\alpha^3 \left({c_s H\over \eta}\right)\left({P\over P_{gas}}\right)^2 >\sim
16,
\end{equation}
and
\begin{equation}
120< \sim\alpha^2 \left({c_sH\over\nu}\right)\left({P\over P_{gas}}\right),
\end{equation}
where $P$ is the total pressure, and $P_{gas}$ is the gas pressure.
{}For temperatures and Mach numbers appropriate for
ionized disks, both of these conditions are usually satisfied.
In this regime the typical flux tube radius is (\cite{v95a}, \cite{v95b})
\begin{equation}
r_t\sim \alpha^{3/2} H{P\over P_{gas}}.
\label{eq:rt}
\end{equation}
The factor of $P/P_{gas}$ enters here because turbulent pumping is
ineffective for photons in accretion disks.  Consequently, the magnetic
pressure in a flux tube  is
limited by the ambient gas pressure instead of the total pressure.

In order to calculate the effects of radial buoyancy, we need to reconsider
the third term in equation (\ref{eq:force}).  The disk properties will
vary radially, and the flux tube's length and radius will change as it
moves.  We need to calculate the change in energy associated with moving
a flux tube radially.
If we consider an empty flux tube of radius $r_t$ and length  $L$
embedded in an accretion disk with an ambient pressure $P$
a length $L$ then it will displace a volume of fluid equal to
its own volume.  Removing (adding) such a flux tube from an accretion disk
will cause the surrounding gas to expand (contract) and lead to a drop
(rise) in the local energy due to pressure work.  The energy involved is
\begin{equation}
U_{tube}=\Delta\left[\int_{-\infty}^{\infty} ({1\over 2}\rho z^2\Omega^2+
{1\over\gamma-1}P)dz\right].
\end{equation}
If we assume that the gas expands adiabatically to fill the volume of the
flux tube as it moves, then we can use the virial theorem for a pressure
supported thin accretion disk, i.e.
\begin{equation}
\int_{-\infty}^{\infty}\rho z^2\Omega^2 dz=\int_{-\infty}^{\infty}P dz,
\end{equation}
to show that
\begin{equation}
U_{tube}=\int_{-\infty}^\infty P(z)\pi r_t^2(z){dL\over dz}dz\sim P(\pi
r_t^2)L,
\end{equation}
where we take the pressure and flux tube radius to be the appropriate
values near the midplane of the disk.  In other words, we obtain the
usual expression for the energy loss due to adiabatic expansion.
The contribution from the gravitational energy term, which is
of comparable order, is canceled by an additional thermal energy term.
This approach fails for a flux tube
in a radiation pressure dominated environment because only the gas
pressure contributes to the energy expended by the gas expanding into
the volume previously occupied by the flux tube.  On the other hand,
the gravitational potential term depends only on the matter density,
and is not reduced by the tendency of the photons to fill up the
otherwise empty flux tube.  Consequently, $U_{tube}$ is of order
$Pr_t^2L$ rather than $P_{gas} r_t^2L$.

Moving a flux tube radially
implies a change in $U_{tube}$ and a corresponding radial buoyancy.
The radial force on a flux tube is equal to the radial
derivative of $U_{tube}$.  Since a given flux tube can split, or
combine with other flux tubes, as it moves outward we need to
constrain the radial variation of $U_{tube}$ by using the conservation
of magnetic flux, i.e. $B_t r_t^2\propto P_{gas}^{1/2}r_t^2$ is a
constant of motion.  Consequently, the radial force per unit length
due to buoyancy is
\begin{equation}
{}F_r\approx -P (\pi r_t^2)\partial_r \ln (P P_{gas}^{-1/2}L).
\end{equation}
Since the magnetic flux tube threads a turbulent disk, its actual
length is $\sim H (H/L_T)\sim H\alpha^{-1/2}$.  This implies
that
\begin{equation}
{}F_r\approx -P(\pi r_t^2)\partial_r {1\over2}\ln (P^2H^2P_{gas}^{-1}
\alpha^{-1}).
\end{equation}
The sign of this radial derivative is not obvious, and may not even be
the same for all radii and at all times.  For a gas pressure dominated
disk we have $P=P_{gas}$ and $\dot M\approx \alpha PH\Omega^{-1}$.
Consequently,
\begin{equation}
\partial_r{1\over 2}\ln (P^2H^2P_{gas}^{-1})=\partial_r{1\over2}\ln(\dot M
c_s\alpha^{-2}),
\end{equation}
and the buoyant force will point outward in a stationary disk.  On the
other hand, near a thermal transition front, with $\partial_r T$ and
$\partial_r\dot M$ negative, magnetic flux lines can be pulled inward.

{}For a radiation pressure dominated disk
$\dot M\approx \alpha PH\Omega^{-1}$, $P_{gas}\propto\Sigma H^{-1}P^{1/4}$,
and $H\propto\dot M$, so that
\begin{equation}
{P^2H^2\over P_{gas}\alpha}\propto T^7\dot M^3\Omega.
\end{equation}
{}For a stationary disk this increases inward, so that once again
we obtain an outward buoyant force on the magnetic field lines.

If we assume that the curvature of a flux tube across the disk is
negligible (more on that later) we obtain the mean radial motion
due to buoyancy by equating the radial buoyant force given above
with the turbulent drag due to a systematic outward motion of a
flux tube.  Averaging the turbulent drag on the flux tube over
many eddies we find that
\begin{equation}
\langle|\vec V_T-\vec V_t|(\vec V_T-\vec V_t)\rangle\approx
-{3\over2}\langle|\vec V_T-\vec V_t|\rangle\vec V_b\sim -|V_T|\vec V_b,
\end{equation}
where we have assumed that $\langle\vec V_T\rangle=0$ and
$\langle \vec V_t\rangle=\vec V_b$.
Dropping constants of
order unity we  obtain
\begin{equation}
{F_r\over L}\approx \rho |V_T| V_r r_t\sim {U_{tube}\over r}.
\end{equation}

This implies that
\begin{equation}
V_r\sim {P\over P_{gas}}\alpha {c_s^2\over r\Omega}.
\label{eq:drift}
\end{equation}
In other words, the radial buoyancy of the field lines gives rise to
an outward drift which is of order the inward drift of
the matter times $P/P_{gas}$.  This same procedure (balancing
buoyancy with turbulent drag) was used in
Vishniac (1995b) to obtain a vertical buoyant velocity of $\sim\alpha c_s$.
Note that we have ignored the specific angular momentum
of the gas entrained on the field lines.  This is justified as long
as the flux tubes are largely empty, as they are in ionized accretion disks.
In a disk dominated by gas pressure, the inward moving matter
will act to drag the field lines
inward, and the actual motion of the field lines will be determined
by the detailed balance between these two effects.  However, in a radiation
pressure dominated environment the field lines will usually move
outward.  This conclusion will be affected by viscous and thermal
instabilities in AGN disks, which may temporarily produce
radial gradients which will pull the magnetic field lines
inward.

When the flux tubes are not empty they will exchange matter, and
consequently angular momentum, with their environment at the
eddy turn over rate, which is $\sim \Omega$.  Neglecting other
effects which will tend to equalize the specific angular momentum
of material inside and outside flux tubes, we see that this
exchange rate implies that the average specific angular momentum of the
flux tubes can differ from the surrounding gas by as much as
\begin{equation}
\Delta(r^2\Omega_{tube})\sim {V_r\over\Omega}\partial_r(r^2\Omega_{gas}).
\end{equation}
Of course, there will be large fluctuations in this quantity, since the
magnetic field will drive an instability by creating pockets of
gas and field with specific angular momenta both larger and smaller than
the local average.  This systematic bias in the specific angular
momentum of gas in the flux tubes will lead to a radial force
of $\sim -\rho_t\Omega V_r$.  Comparing this to the buoyant force
per unit volume and using equation (\ref{eq:drift}) we have
\begin{equation}
{-\rho_t V_r\Omega\over (\rho-\rho_t) c_s^2/r}\sim {\rho_t\over \rho-\rho_t}
{P\over P_{gas}}\alpha.
\end{equation}
Since the radiation pressure differential across a flux tube is
negligibly small (\cite{v95b}) the magnetic pressure in a tube
must be balanced by a drop in the matter pressure.  Therefore,
\begin{equation}
\alpha\sim {\langle B^2\rangle\over 8\pi P}<{B_t^2\over 8\pi P}
\sim{\Delta P_{gas}\over P}\sim {(\rho-\rho_t)\over\rho} {P_{gas}\over P}.
\end{equation}
{}From this we conclude that the ratio of the forces due to the specific
angular momentum of material inside flux tubes to the buoyant forces
will be less than $\rho_t/\rho$, which is less than one.  Of course,
this limit will be approached only when the flux tubes are barely
distinguishable from the rest of the fluid, so that the volume average
magnetic pressure is close to the magnetic pressure in the flux tubes.
If we can neglect specific angular momentum, then the radial velocity
due to buoyancy depends on the flux tube radius and internal magnetic
field only in the combination $B_t^2 r_t$, which is always of order
$V_T^2 L_T$ for a magnetic field embedded in a turbulent medium and
in equipartition with the turbulent energy (\cite{v95a}).
We conclude that the results presented here should apply to all accretion
disks except for those where the resistivity is so large as to nearly
obliterate all flux tube structure.

Up to now we have assumed that the magnetic field outside the accretion
disk crosses the disk with a negligible bending angle.  This will not be
the case if the external field is driving a magneto-centrifugal wind
(e.g. \cite{bp82}).  In this case the motion of the field can determined
by vertical mixing through the disk. We can see this effect for a weak
external field by using the usual mean field theory, and ignoring the
division of the magnetic field in the disk into flux tubes.
Our argument is equivalent to the one given by Van Ballegooijen
(1989).  The only conceptual difference is that we have justified the
concept of turbulent diffusion by appealing to a specific physical
model for the microscopic structure of the magnetic field.
Starting from the induction equation
\begin{equation}
\partial_t\vec B=\vec\nabla\times\vec v\times\vec B,
\end{equation}
we can define a transient field $\vec b$ which is the response of
the large scale magnetic field, $\vec B_0$, to small scale turbulent motions.
We have
\begin{equation}
\partial_t\vec b=\vec\nabla\times\vec v\times\vec B_0.
\end{equation}
Substituting this back into the right hand side of the induction equation
we obtain
\begin{equation}
\partial_t\vec B_0=\vec\nabla\times\langle \vec v\times
(\int ^t \vec v\cdot\vec\nabla B_0 dt)\rangle -\vec \nabla\times
\langle \vec v\times (\int^t \vec v\cdot\vec\nabla \vec B dt)\rangle,
\end{equation}
for incompressible turbulence.  The first term in this equation is
the source of the turbulent dynamo, if any, and the second term gives
rise to diffusive effects.  If we consider only the diffusive terms which
affect the evolution of $B_z$ then we get
\begin{equation}
\partial_t B_z = -\partial_r\left(\langle v_z^2\tau\rangle\partial_z B_r
\right)
+\partial_r\left(\langle v_r^2\tau\rangle\partial_rB_z\right)
+\partial_r \left(\langle v_z v_r\tau\rangle (\partial_z B_z-\partial_rB_r)
\right),
\label{eq:diffuse}
\end{equation}
where $\tau$ is the correlation time of the turbulence, in this case
$\sim \Omega^{-1}$. The last term will be small, not only because the
off diagonal terms of the averaged stress tensor will be small, but
because the extra radial derivative of $B_r$ will make it of order $H/r$
compared to the first term. (Since the divergence of the mean magnetic
field vanishes, the two parts of the last term are equal to each other.)
This implies an outward drift of the
vertical field lines given by
\begin{equation}
V_r= \langle v_z^2\tau\rangle {\partial_z B_r\over B_z}\sim \alpha
c_s\tan\theta,
\end{equation}
where $\theta$ is the bending angle of the magnetic field as it passes
through the disk.  Similar, and more precise, estimates can be
found in paper by Lubow et al. (1994) and Reyes-Ruiz \& Stepinski (1996).
A different result can be obtained by assuming that the
level of turbulent diffusivity in the disk
is much less than that required to explain radial angular momentum transport.
This is conceivable only if the Alfv\'en velocity in the disk is
supersonic, thereby suppressing the Balbus-Hawley mechanism.
If the magnetic field is driving a wind, then
$\theta$ will be positive and of order unity and this will dominate over
the radial
buoyancy of field lines for $(P_{gas}/P)>(H/r)$.  On the other hand,
if the field's geometry is not directly affected by the disk, and
has a curvature scale comparable to $r$ then $\theta$ will be of order
$H/r$ and the resulting radial drift velocity  will be of the same
order as the inward velocity of the matter.  In this case
radial buoyancy effects will dominate for radiation pressure
dominated disks.  Finally, in the event that the magnetic field lines
bend {\it inward} as they cross the disk, this diffusive effect will
move the vertical field lines inward.

We note that this derivation neglects
the tension due to the bending of the large scale field as it crosses
the disk and the possibility that the angular momentum in the disk is
transported out into a magnetically driven wind instead of radially
within the disk.  Both assumptions are justified for a weak external
field.  A precise estimate of when this will run into problems must
necessarily await a consensus on the degree to which an external
field can produce a torque.  Successful wind and jet models
based on an external poloidal field (\cite{bp82}, \cite{NS94},
\cite{SNRL94}, and \cite{SNORWL94}) give an equivalent `$\alpha$'
based on an external torque which can be as much as
\begin{equation}
\alpha_{ext}\sim {B_{ext}^2\over P} {r\over H},
\label{eq:ext}
\end{equation}
but these models assume that large radial
bending angles near the disk
are possible.  As mentioned above, this is reasonable only for
an induced magnetic pressure in the disk which is at least as large
as the ambient thermal pressure.  At smaller values the external
field will induce turbulence, and transport, within the disk.

Without a definite model for the external torque, it is
difficult to compare external and internal torques, but
there are two points to consider with respect to the relationship
between the internal turbulent viscosity and $B_{ext}$.
First, the formation of flux tubes implies that the same amount
of flux is compressed into a smaller volume, giving rise to
a higher internal rms $B_z$ than $B_{ext}$.  If the flux tube
formation process is very efficient, as would expect in an
ionized disk, then
\begin{equation}
B_{z,internal}\sim B_{z,ext}^{1/2}(8\pi P_{gas})^{1/4}.
\end{equation}
Moreover, the $\alpha_{int}$
associated
with an imposed vertical field is $\sim (V_{Az}/c_s)$ rather than
$\sim(V_{Az}/c_s)^2$ (\cite{vd93}).  Consequently, we have
\begin{equation}
\alpha_{int}> \left({B_{z,ext}\over \sqrt{4\pi\rho} c_s}\right)^{1/2}.
\end{equation}
Of course, this limit
neglects any magnetic field generated within the disk itself, but
such a field will only increase $\alpha_{int}$.  We see that in
spite of the potential advantage inherent in external torques
due to their extended moment arm, reflected in the factor of
$r/H$ in equation (\ref{eq:ext}), internal torques are likely
to dominate for $V_{Az}<c_s$.

{}Finally, we note that the second term in equation (\ref{eq:diffuse}) will
tend to smooth
out radial gradients in $B_z$.  When the scale length for such gradients
is of order $r$, then this will move vertical magnetic flux radially
at a speed comparable to the accretion velocity.  This would prevent
radial infall from building up the vertical magnetic field at small
radii to arbitrarily high values even in the absence of radial buoyancy.
In gas pressure dominated disks it also prevents radial buoyancy from
expelling vertical flux from the disk, even in the absence of a disk
dynamo.  In a radiation pressure dominated disk this effect will be smaller,
and will allow a positive $\partial_r\ln B_z$ of order $(P /P_{gas})r^{-1}$.

We conclude that the strong magnetic fields required for nonthermal emission
from the central parts of AGN disks must be either be maintained by
dynamo processes at small radii, or dragged in through a magnetically
dominated, and relatively nonturbulent, outer disk.  In the latter case,
the vertical field will adjust to changes in the disk
at a the characteristic rate of
\begin{equation}
\tau_B^{-1}={V_r\over r}\sim \alpha {H\over r}\Omega.
\end{equation}
The former case will allow the field to adjust on a dynamo time
scale, which will be (\cite{v95b})
\begin{equation}
\tau_{dynamo}^{-1}\sim \left({P\over P_{gas}}\right)\alpha
\end{equation}

The nature of the dynamo at small radii is beyond the scope of this
paper.  Most work on disk dynamos has concentrated on the generation
of toroidal fields, but the same processes will produce a poloidal
field, albeit at a much reduced rate.  In this case the field lines will
tend to open up as differential rotation pumps energy into the
disk magnetosphere, ultimately producing a large scale poloidal field.

\section{Implications for the Variability of AGN}

We have examined the radial buoyancy of poloidal field lines
embedded in an accretion disk.  In general the field lines will
move outward through buoyant forces at a rate comparable to the
inward drift caused by accretion.  This leaves the question of their
global evolution
unsettled.  However, in radiation pressure dominated disks radial
buoyancy dominates over accretion unless the magnetic field
is strong enough to suppress turbulence.  It follows that
the global magnetic field embedded in accretion
disks around AGN is likely to have its origin in dynamo processes taking
place at small radii.  In reaching this conclusion we have assumed
that the angular momentum transport in these disks is mediated by
magnetic field instabilities.  Since the enhanced vertical buoyancy
of magnetic field lines in such disks will tend to decrease the
efficiency of this process (\cite{v95b}) it seems possible that
this assumption may be violated in the presence of hydrodynamic
transport processes.  As an example, we cite the possible role of
internal waves, which normally give $\alpha\sim (H/r)^2$ (\cite{vd89}),
a dimensionless viscosity which is unlikely to dominate in disks
where the radiation pressure is negligible.

What does this imply about the variability of nonthermal radiation
from AGNs? If the poloidal field is generated from within the
disk, and moves outward through turbulent reconnection and radial
buoyancy, then
both its generation rate and its outward migration rate will depend on
conditions within the disk. Consequently, variations in the state of
the disk will lead to changes in the basic parameters
of the magnetosphere, which will lead directly to variations in
its nonthermal radiation. Variations in the mass accretion rate
in the inner disk will arise
from instabilities operating at larger radii and will reflect the
thermal rates at those radii, roughly $\alpha \Omega(r)$.
Since the time scales for disk variations
will approach the light travel time across the magnetosphere for
the inner regions of disks accreting near the Eddington limit it
will be necessary to include time derivatives in the formulae for the
structure of the magnetosphere.

As long as the poloidal field has only a weak effect on the mass
flow within the disk, compared to internal processes driving
angular momentum flow, the effects discussed here are incapable of
driving disk instabilities.  However, if the radiation pressure
dominated part of the disk is subject to local instabilities over
its entire radius, then such instabilities will lead directly
to variations in the structure of the magnetosphere and the
nonthermal emission it drives.  For example, thermal transition
fronts will produce narrow annuli with a large local variations
in $\partial_r \dot M$, driving the magnetic flux in the direction
of the local mass flow.
Since the disk instabilities will
reflect the thermal and mass transport time scales for a wide
range of radii, from the inner edge of the disk out to the radius
where the radiation pressure of the disk loses its dominance,
these time scales will also be present in the nonthermal emission.
The upper limit to these time scales will depend on the details
of the disk structure, but will clearly be much longer than
orbital periods in the inner part of the disk.  However, the
shortest time scales for variations driven in this way will be
comparable to the period given in Eq. (\ref{eq:1p1}).

\acknowledgements
This work has been supported in part by NASA grants
NAGW-2418 and NAG5-2773 (ETV) and by the Basic Research
Project 95-5200-002 of the Ministry of Science and Technology,
Korea (SJP).


\begin{thebibliography}{}
\bibitem[Balbus \& Hawley\ 1991]{bh91} Balbus, S., \& Hawley,
J.F.\ 1991, \apj, 376, 214
\bibitem[Blandford \& Payne\ 1982]{bp82}Blandford, R.D., \&
Payne, D.G. \ 1982, \mnras, 199, 883
\bibitem[Blandford \& Znajek\ 1977]{bz77}Blandford, R.D., \&
Znajek, R.L.\ 1977, \mnras, 179, 433
\bibitem[Chandrasekhar \ 1960]{c60}Chandrasekhar, S.\ 1961,
Hydrodynamic and Hydromagnetic Stability, Oxford University Press: London
\bibitem[Contopoulos\ 1995]{C95}Contopoulos, J., \apj, 450,
616
\bibitem[Lightman \& Eardley\ 1974]{le74}Lightman, A.P.
\& Eardley, D.M.\ 1974, \apjl, 187, 897
\bibitem[Lubow, Papaloizou, \& Pringle\ 1994]{LPP94}Lubow, S.H.,
Papaloizou, J.C.B., \& Pringle, J.E.\ 1994, \mnras, 267, 235
\bibitem[Macdonald \& Thorne\ 1982]{mt82} Macdonald, D. A., \&
Thorne, K. S.\ 1982, \mnras, 198, 345
\bibitem[Najita \& Shu\ 1994]{NS94}Najita, J.R., \& Shu, F.H.\ 1994,
\apj, 429, 808
\bibitem[Park \& Vishniac\ 1989a]{pv89a}Park, S.J., \&
Vishniac, E.T.\ 1989a, \apj, 337, 78
\bibitem[Park \& Vishniac\ 1989b]{pv89b}Park, S.J., \&
Vishniac, E.T.\ 1989b, \apj, 347, 684
\bibitem[Reyes-Ruiz \& Stepinski\ 1996]{RP96}Reyes-Ruiz, M.,
\& Stepinski, T.F.\ 1996, \apj, 459, 663
\bibitem[Shu, Najita, Ruden, \& Lizano\ 1994]{SNRL94}Shu, F.H.,
Najita, J.R., Ruden, S.P., \& Lizano, S.\ 1994, \apj, 429, 797
\bibitem[Shu, Najita, Ostriker, Ruden, Wilken, \& Lizano\ 1994]{SNORWL94}
Shu, F.H., Najita, J.R., Ostriker, E.C., Ruden, S.P., Wilken, F.,
\& Lizano, S.\ 1994, \apj, 429, 781
\bibitem[Thorne \& Macdonald\ 1982]{tm82} Thorne, K. S., \&
Macdonald, D. A.\ 1982,
\mnras, 198, 339
\bibitem[Thorne, Price \& Macdonald\ 1986]{tpm86}Thorne, K. S.,
Price, R. H.,
\& Macdonald, D. A.\ 1986, Black Holes: The Membrane Paradigm, (Yale University
Press: New Haven)
\bibitem[Van Ballegooijen\ 1989]{vb89}Van Ballegooijen, A.A.\
1989, in Accretion Disks and Magnetic Fields in Astrophysics,
(Kluwer: Dordrecht) p. 99
\bibitem[Velikhov\ 1959]{v59}Velikhov, E.P.\ 1959, Soviet
JETP, 35, 1398
\bibitem[Vishniac\ 1995a]{v95a} Vishniac, E.T.\ 1995a, \apj, 446,
724
\bibitem[Vishniac\ 1995b]{v95b} Vishniac, E.T.\ 1995b, \apj, 451,
816
\bibitem[Vishniac \& Diamond\ 1989]{vd89}Vishniac, E.T.,
\& Diamond, P.H.\ 1989, \apj, 347, 435
\bibitem[Vishniac \& Diamond\ 1992]{vd92}Vishniac, E.T., \&
Diamond, P.H.\ 1992, \apj, 398, 561
\bibitem[Vishniac \& Diamond\ 1993]{vd93}Vishniac, E.T., \&
Diamond, P.H.\ 1993, in Accretion Disks in Compact Stellar Systems,
ed. J.C. Wheeler (World Scientific Press: Singapore), p. 41
\end{thebibliography}
\end{document}